\documentclass[twocolumn, twocolappendix]{aastex631}
\usepackage{xcolor}
\usepackage{xspace}
\usepackage{amsmath}
\usepackage{CJKutf8}
\usepackage{threeparttable}
\usepackage{tabularx}
\usepackage{makecell}
\usepackage{multirow}
\usepackage{orcidlink}

\begin{document}
\newcommand{\hatb}[1]{\hat{\boldsymbol{{#1}}}}
\newcommand{\rev}[1]{\color{orange}#1 \color{black}}

\begin{CJK*}{UTF8}{gbsn}
\title{A third star in the HAT-P-7 system, and a new dynamical pathway to misaligned hot Jupiters}
\author{Eritas Yang (杨晴) \orcidlink{0009-0005-2641-1531}}
\affiliation{Department of Astrophysical Sciences, Princeton University, 4 Ivy Lane, Princeton, NJ 08540, USA}
\author{Yubo Su (苏宇博) \orcidlink{0000-0001-8283-3425}}
\affiliation{Department of Astrophysical Sciences, Princeton University, 4 Ivy Lane, Princeton, NJ 08540, USA}
\author{Joshua N. Winn (温乔书) \orcidlink{0000-0002-4265-047X}}
\affiliation{Department of Astrophysical Sciences, Princeton University, 4 Ivy Lane, Princeton, NJ 08540, USA}

\begin{abstract}
The retrograde orbit of the hot Jupiter HAT-P-7b is suggestive of high-eccentricity migration caused by dynamical interactions with a massive companion. However, the only other known body in the system is an M dwarf located $\sim$10$^3$~AU away, too distant to cause high-eccentricity migration without fine tuning. Here we present transit-timing and radial-velocity evidence for an additional stellar companion with semi-major axis $32^{+16}_{-11}$~AU, eccentricity $0.76^{+0.12}_{-0.26}$, and minimum mass $0.19^{+0.11}_{-0.06}$~$\rm M_\odot$. 
We investigate several dynamical routes by which this 
nearby companion star could have played a role in converting a cold Jupiter into the retrograde hot Jupiter that is observed today. Of particular interest is a novel ``eccentricity cascade'' mechanism involving both of the companion stars: the outer companion periodically excites the eccentricity of the inner companion through von Zeipel–Lidov–Kozai (ZLK) cycles, and this eccentricity excitation is slowly transferred to the cold Jupiter via successive close encounters, eventually triggering its high-eccentricity migration. The plausibility of this mechanism in explaining HAT-P-7b shows that stellar companions traditionally considered too distant to cause hot Jupiter formation might nevertheless be responsible, with the aid of closer-orbiting massive companions. With these developments, HAT-P-7b is one of the few hot Jupiters for which a complete high-eccentricity migration history can be simulated based only on observed bodies, rather than invoking bodies that are beneath detection limits or that are no longer in the system.

~\\ 
\end{abstract}

\section{Introduction}
\end{CJK*}

HAT-P-7b is a hot Jupiter with a mass of $1.8\,\rm M_J$ and a radius of $1.4\,\rm R_J$ orbiting an F6V star ($M_\star = 1.35\,\rm M_{\odot}$) with a period of $2.2\,\rm days$ \citep{Pal2008}. The host star was one of the first known to be grossly misaligned with the orbit of a planet, originally through observations of the Rossiter-McLaughlin effect ($\psi=94.6^\circ$$^{+5.5}_{-3.0}$, \citealp{Winn2009}; $\psi>85.7^\circ$, \citealp{Narita2009}), and later via asteroseismology ($\psi\approx120^\circ$, \citealp{Benomar2014}; $83^\circ<\psi<111^\circ$, \citealp{Lund2014}) and the gravity-darkening method ($\psi=101^\circ\pm2^\circ$ or $\psi=87^\circ\pm2^\circ$, \citealp{Masuda2015}), all of which indicate a retrograde or nearly polar orbit. 

Spin-orbit misalignments are common among hot Jupiters \citep{Triaud2010, Albrecht2022} and are often interpreted as evidence for a high-eccentricity (high-e) migration process \citep{Hebrard2008, Winn2010, Dawson2018, Rice2022}. In such processes, a cold Jupiter's orbital eccentricity is increased enough for tidal effects to gradually shrink and circularize the orbit. The eccentricity excitation might also be accompanied by inclination excitation, thereby explaining the spin-orbit misalignment. Some examples of high-e migration scenarios include secular chaos \citep{Wu2011, Hamers2017, Teyssandier2019}, planet-planet scattering \citep{Rasio1996, Beauge2012}, von-Zeipel-Lidov-Kozai (ZLK) oscillations induced by planetary companions \citep{Naoz2011, Dawson2014, Petrovich2016} or stellar companions \citep{Wu2003, Fabrycky2007, Petrovich2015, Anderson2016, Vick2019, vicksulai2023}, and a combination of planet-planet scattering and ZLK migration \citep{Nagasawa2008, Lu2025}.

The scenario that appears most likely to explain misalignments as drastic as seen in HAT-P-7 is ZLK migration, particularly when it is induced by stellar companions \citep[see, e.g.,][]{Fabrycky2007,Albrecht2022}.
However, searches for distant stellar companions to the hosts of hot Jupiters have concluded that only $\sim 10\%$ of hot Jupiters could have migrated via ZLK oscillations induced by a stellar companion \citep{Wu2007, Ngo2016}. In most cases, the observed stellar companions are either too distant or too low in mass to induce significant ZLK oscillations \citep{Ngo2016}.

The HAT-P-7 system illustrates this point. The host star has a directly imaged companion star with
estimated spectral type M5.5V and a projected separation of 1240~AU \citep{Narita2012}. For ZLK-driven migration to be effective in systems with such distant companions, the companion’s initial orbital plane must be finely tuned to be almost perpendicular to that of the planet \citep{Wu2003, Gupta2024}. This stringent geometric requirement disfavors such a migration pathway for HAT-P-7b and similar systems.

On the other hand, the HAT-P-7 host star also shows a long-term radial velocity trend that is indicative of an additional massive companion \citep{Winn2009, Narita2012}. However, no firm constraints on this inner companion's mass or orbit have been obtained to date.
Below, we show that this additional companion is an M dwarf on a highly eccentric
orbit with a semi-major axis smaller than $\sim$70~AU. The evidence is presented in Section~\ref{sec:data}.
Then, in Section~\ref{sec:dynamics}, we take advantage of an improved knowledge of the inner companion's properties to investigate potential dynamical pathways that could have
converted a giant planet on a wide circular orbit into a retrograde-orbiting hot Jupiter resembling HAT-P-7b. We summarize our results and discuss their implications in Section~\ref{sec:discussion}. Hereafter, we refer to quantities relating to the hot Jupiter with subscript “p” (for planet). We use the subscript ``1'' to refer to the newly characterized inner stellar companion, and ``2'' to refer to the directly imaged outer stellar companion.\\

\section{Data\label{sec:data}}

\subsection{Radial velocities}

\cite{Winn2009} first presented evidence for the inner companion based on radial velocity (RV) data collected between 2007 and 2009. They reported a radial acceleration of $\dot{\gamma} = 21.5\pm2.6\, \rm m/s/yr$, consistent with the subsequent measurement by \cite{Narita2012} of $\dot{\gamma} = 20.3\pm1.8\, \rm m/s/yr$ using data from 2008 to 2010. Since the available RV data were consistent with a constant acceleration over this entire time period, little information could be obtained about the companion's orbit beyond a lower limit of $\sim$10~years on the orbital period and an order-of-magnitude constraint on $M_1/a_1^2$.

\begin{figure*}
    \centering
    \includegraphics[width=0.73\linewidth]{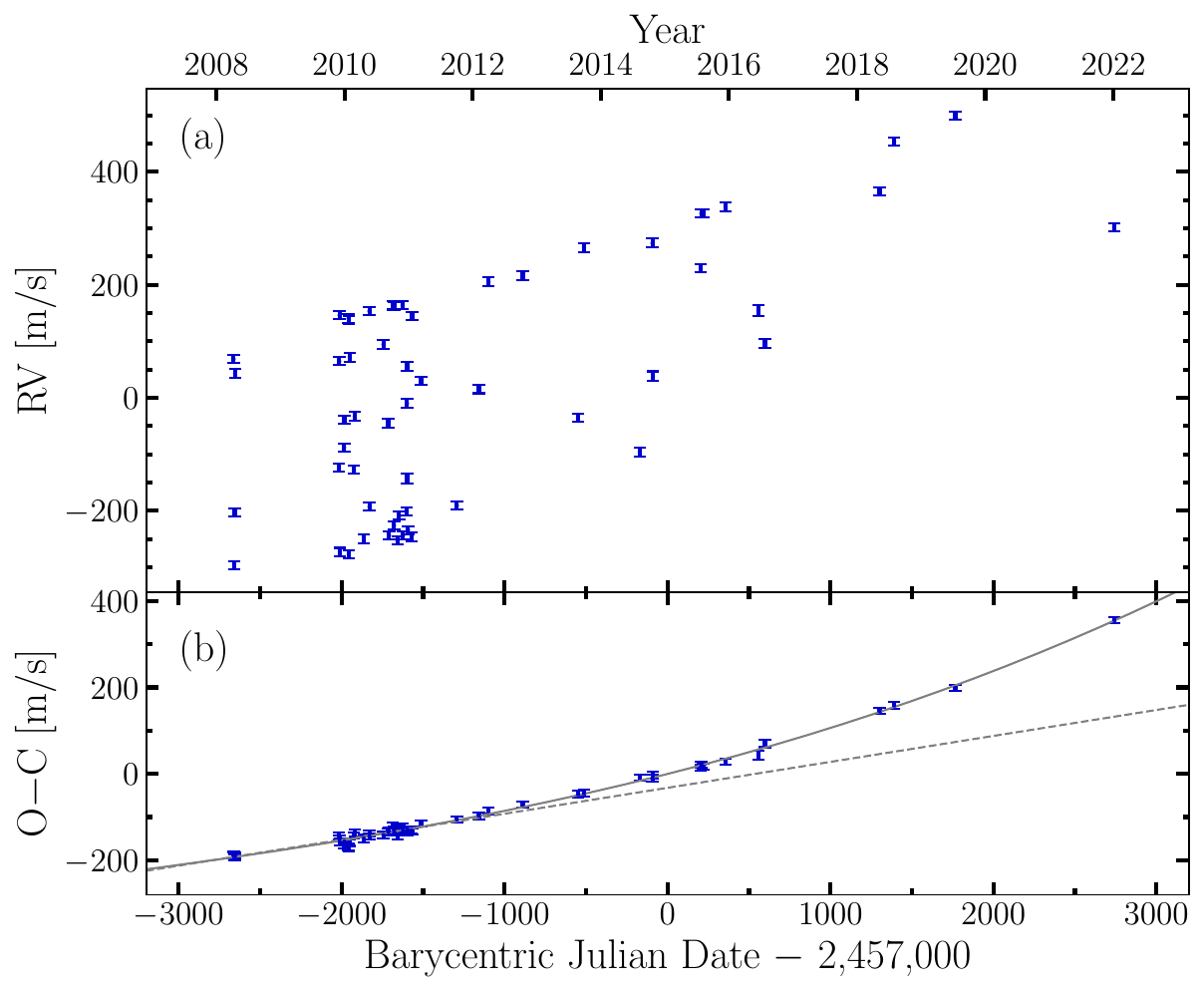}
    \caption{(a) Radial velocity variation of HAT-P-7 observed with Keck/HIRES from 2007 to 2022. (b) Residuals after subtracting the best-fit sinusoidal model. The residual trend is well-fit by a cubic function of time (solid line). The dashed line represents a linear fit to the data between 2007 and 2010, and is consistent with the radial accelerations measured by \cite{Winn2009} and \cite{Narita2010}.}
    \label{fig:RV}
\end{figure*}

Figure~\ref{fig:RV} presents 15 years of RV measurements using the High Resolution Echelle Spectrometer (HIRES; \citealp{Vogt1994}) on the Keck~I telescope on Mauna Kea, Hawaii. Most of these data were published previously by \cite{Knutson2014} and \cite{Wong2016}; to these we added four data points obtained between 2018 and 2022. The top panel shows the RVs and the bottom panel shows the residuals after subtracting the best-fit sinusoidal model, representing the contribution from the hot Jupiter. 
As a preliminary description of the long-term trend, we fitted the residuals with a cubic function of time, finding 
\begin{equation}
\begin{aligned}
    v_{r,\rm res}~ =& \left(34.9\pm0.7 \,\rm \frac{m/s}{yr}\right) t + \left(1.40\pm0.09 \,\rm \frac{m/s}{yr^2}\right) t^2 \\
    &+ \left(0.033\pm0.018 \,\rm \frac{m/s}{yr^3}\right) t^3,
\end{aligned}
\end{equation}
where $t\equiv {\rm BJD} - 2{,}457{,}000$.
All three coefficients are positive, whereas a purely sinusoidal function of time
would have first and third coefficients of opposite signs. Thus, the results are suggestive of a companion on an eccentric orbit.

\subsection{Transit timing variations}
As the center of mass of the HAT-P-7 system accelerates away from the observer, the measured time interval between successive transits will steadily grow, due to the R{\o}mer effect. To verify this prediction and to gain additional leverage on the parameters of the companion responsible for the radial-velocity trend, we analyzed the transit timing variation (TTV) data spanning over a decade.

The transit timing catalog compiled by \cite{Holczer2016} provides 554 transit times for HAT-P-7b, based on observations between 2009 and 2013 by the Kepler mission \citep{Kepler}. This dataset was derived from PDC-MAP light curves with 30-minute time sampling, and the transit time uncertainties were estimated as (100~min)/(SNR of individual transits). 
In addition, we incorporated more recent observations from the TESS mission \citep{TESS}, which observed 89 transits between 2019 and 2024, and the data are available with 2-minute time sampling. Mid-transit times were determined by fitting each light curve with the standard \cite{Mandel2002} model. We processed the data and extracted transit times and uncertainties following the same procedures described by \cite{Ivshina2022}.

\begin{figure*}
    \centering
    \includegraphics[width=0.75\linewidth]{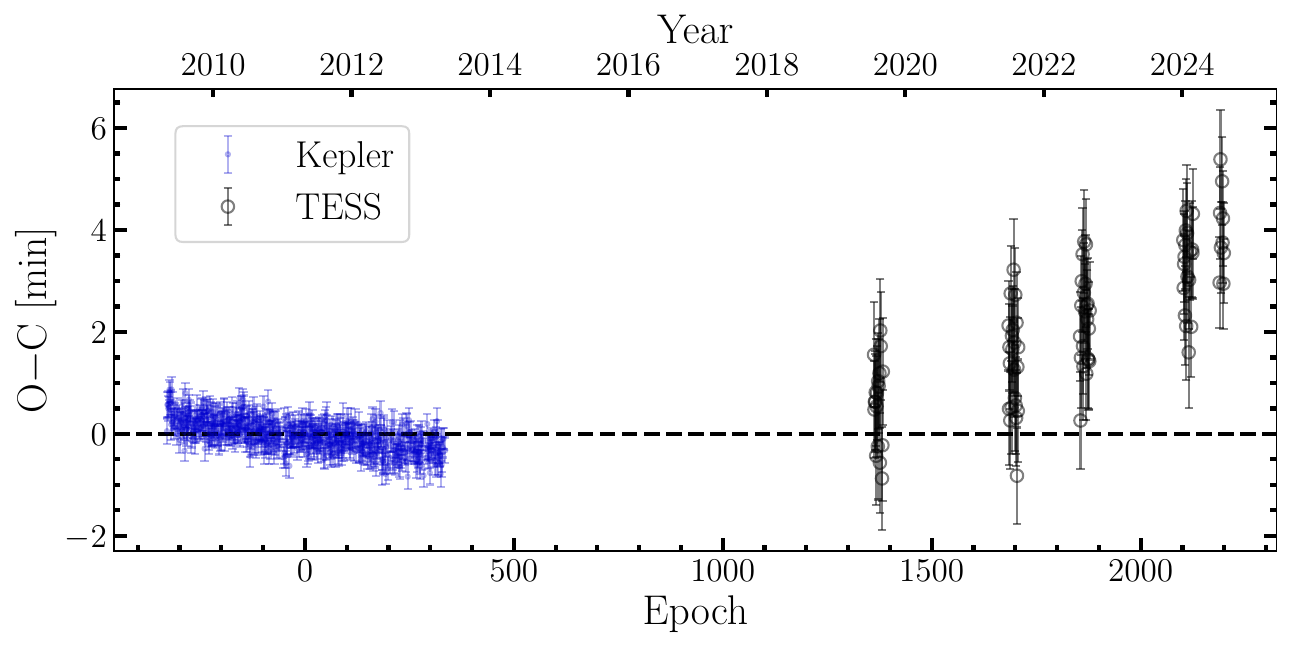}
    \caption{Transit timing residuals between the data and the best-fitting constant-period model.}
    \label{fig:TTV}
\end{figure*}

The TTV data are presented in Figure~\ref{fig:TTV}. The observed trend is consistent with an increasing apparent orbital period, likely caused by the relative acceleration of the HAT-P-7 system.

\subsection{Joint analysis \label{ssec:joint_analysis}}

We fitted the RV and TTV data jointly. The RV time series was modeled in the usual way, as
\begin{equation}
    v_r = K_p\cos(n_pt+\omega_p) + K_1[\cos(f_1+\omega_1) + e_1\cos(\omega_1)] + v_0, \label{eq:RV_fit}
\end{equation}
where $K_p$, $n_p$, $\omega_p$ are the semi-amplitude, mean motion, and argument of pericenter of the hot Jupiter; and
$f_1$, $\omega_1$, $e_1$, and $K_1$
are the true anomaly, argument of pericenter,
eccentricity, and semi-amplitude of the inner
companion.
The constant term $v_0$ accounts for the arbitrary instrumental velocity offset.

The transit-timing model was
\begin{equation}
    t_{\rm mid}(N) = t_0 + P_pN + \frac{1}{c} \int_{t_0}^{t_0+P_p N} v_{r,1}\,dt, \label{eq:TTV_fit}
\end{equation}
where $v_{r,1}$ is the radial velocity variation induced
by the inner companion (the second term in Eq.~\ref{eq:RV_fit}), $P_p$ is the planet's orbital period, and $N$ is the transit epoch.

\begin{figure*}
    \centering
    \includegraphics[width=0.75\linewidth]{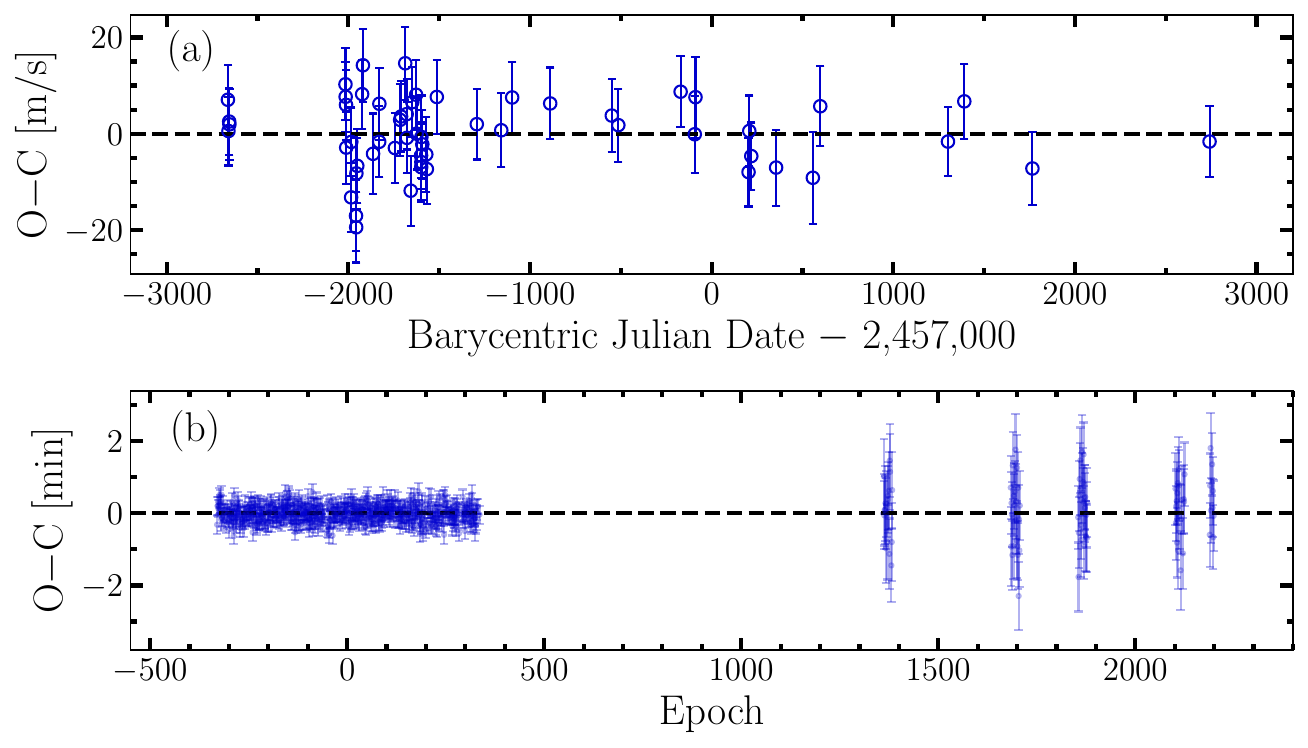}
    \caption{Residuals from the best-fit solution to the RV and TTV data. The combined fit has a $\chi^2$ value of 696 with 689 degrees of freedom. (a) Residuals from the RV fit, which has $\chi^2=52$ and 46 degrees of freedom. (b) Residuals from the TTV fit, which has $\chi^2 = 644$ and 636 degrees of freedom.}
    \label{fig:residuals}
\end{figure*}

\begin{figure}
    \centering
    \includegraphics[width=1\linewidth]{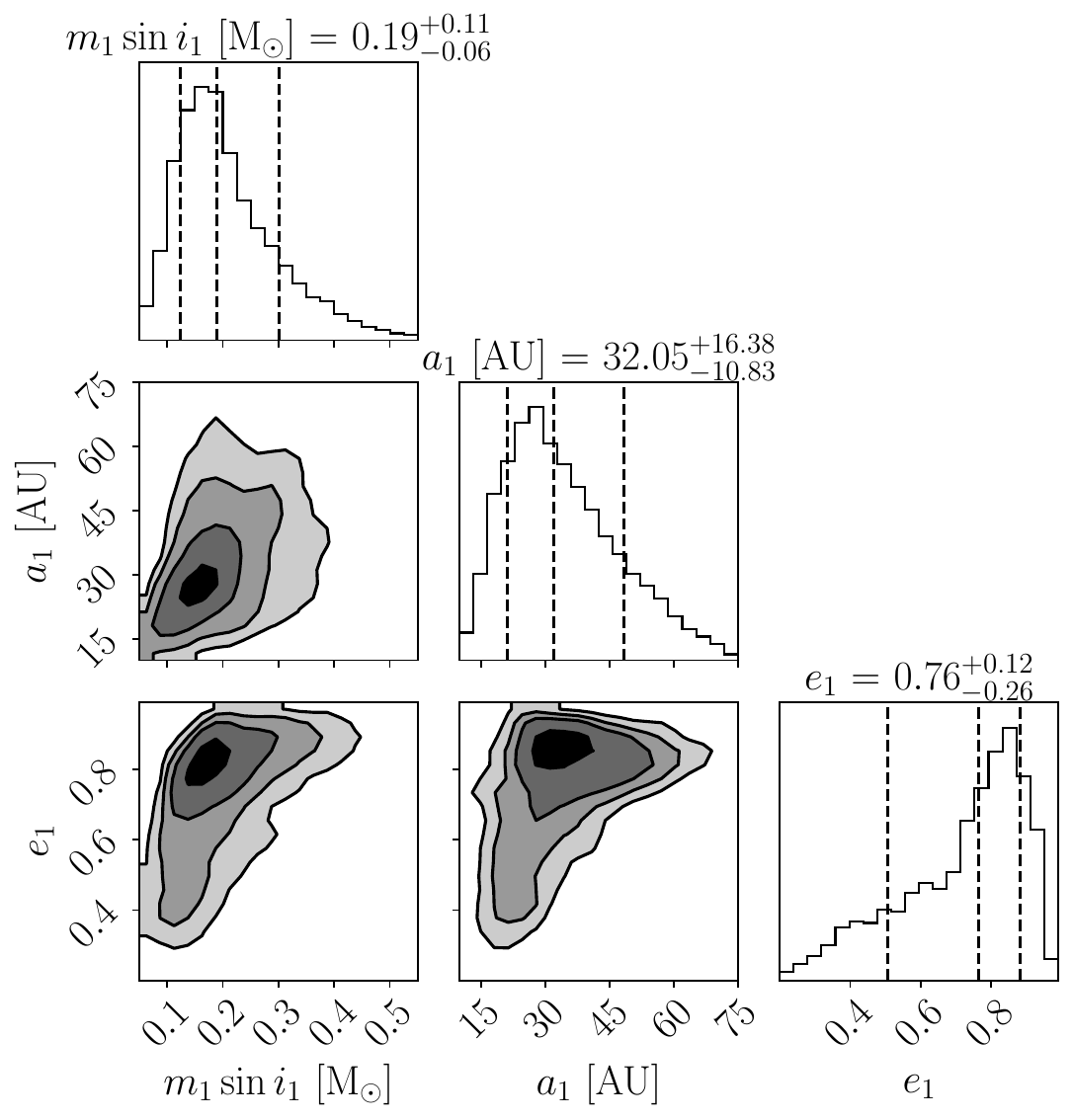}
    \caption{Posterior distributions of the inner companion’s orbital parameters, derived from the RV and TTV data. The contours indicate 0.5, 1, 1.5, and 2-$\sigma$ confidence level, enclosing 11.8\%, 39.3\%, 67.5\%, and 86.4\% of the samples.}
    \label{fig:corner}
\end{figure}

The best-fit values of the parameters and their uncertainties were determined using the \texttt{emcee} package \citep{emcee}. The ensemble sampler was initialized with 192 walkers, and a total of $3.4\times10^5$ samples were collected after discarding the initial $1\times10^5$ burn-in steps. Figure~\ref{fig:residuals} shows residuals from the best-fit solution. The $\chi^2$ value of the best-fit model is 696, and the total number of degrees of freedom of 689.
Figure~\ref{fig:corner} shows the posterior probability distributions for some key parameters, revealing strong covariances
due to the limited observational timespan. The $1$-$\sigma$ constraints on the inner companion's orbital parameters are $m_1\sin{i_1} = 0.19^{+0.11}_{-0.06}\,\rm M_{\odot}$, $a_1=32^{+16}_{-11}$~AU, and $e_1 = 0.76^{+0.12}_{-0.26}$.

\subsection{Astrometric detectability}

The detection of astrometric motion of HAT-P-7 could provide improved constraints on orbital parameters and access to orbital inclinations. For example, \cite{an2024} performed a joint orbital fit of the HAT-P-11 system using radial velocity data and astrometry from the Hipparcos and Gaia missions, allowing them to constrain the orbital orientations of the planets.

To set expectations for astrometric motion in the HAT-P-7 system, we computed the astrometric acceleration of the host star based on the orbital solutions for the inner companion derived from the joint analysis presented in Section~\ref{ssec:joint_analysis}. From the difference in proper motion between epochs 2000 and 2016, we estimated a time-averaged acceleration of $0.017^{+0.015}_{-0.009}$~mas/yr$^{-2}$.

To check on the detectability of
astrometric motion, we examined the
entries for HAT-P-7 in the Tycho-2 catalog \citep{tycho2} and the Gaia DR3 catalog \citep{GaiaDR3};
see Table~\ref{tab:astrometry}.
The tabulated uncertainties
are too large to allow
for a secure detection of an acceleration with the expected amplitude of $\sim$0.02~mas/yr$^{2}$. The Gaia DR3 Renormalized Unit Weight Error (RUWE) of 0.97 is consistent with the absence of proper motion acceleration.
The Gaia DR3 proper
motion differs from the Tycho-2 proper motion by about 2$\sigma$,
and it differs from
the long-term proper motion
implied by the Gaia and Tycho-2
positions by about 3$\sigma$.
These marginal discrepancies,
if real, would imply proper motion
accelerations of order $\sim$0.2~mas/yr$^{2}$, an order of magnitude larger than the
expected signal.

Given the low statistical significance
of the proper motion anomalies,
we did not attempt to interpret
or model them.
Nonetheless, we predict a net change in proper motion of $0.13^{+0.10}_{-0.07}$~mas/yr$^{-1}$ between the beginning and end of the Gaia mission. This signal might be within reach of future Gaia data releases.

\begin{table*}[]
    \centering
    \caption{Astrometric data of the HAT-P-7 host star. The table lists the star’s right ascension ($\alpha$), declination ($\delta$), and proper motions in right ascension ($\mu_\alpha$) and declination ($\mu_\delta$), all referenced to the ICRS frame. Values are taken from the Tycho-2 catalog and the Gaia DR3 catalog.}
    \begin{tabular}{l c c c c c}
    \hline
         & Epoch & mean $\alpha$ (ICRS) & mean $\delta$ (ICRS) & $\mu_\alpha$(mas/yr) & $\mu_\delta$ (mas/yr) \\
    \hline
        Tycho-2 & 2000 & $292.24732138^\circ \pm 27~{\rm mas}$ & $47.96950268^\circ \pm 25~{\rm mas}$ & 
        $-14.8 \pm 1.5$ & $8.7 \pm 1.4$\\
        Gaia DR3 & 2016 & $292.24718620019^\circ \pm 0.010~{\rm mas}$ & $47.96954406410^\circ \pm 0.011~{\rm mas}$ & $-18.325 \pm 0.014$ & $8.851 \pm 0.014$\\
    \hline
    \multicolumn{4}{c}{proper motion inferred from Tycho-Gaia positional shifts} & $-20.4 \pm 1.1$ & $9.3 \pm 1.6$\\
    \hline
    \end{tabular}
    \label{tab:astrometry}
\end{table*}

\section{Dynamical History \label{sec:dynamics}}

Next, we consider the role that the inner companion might have played in sculpting the HAT-P-7 system.
Hot Jupiters are generally not expected to form in situ (\citealp{Rafikov2005, Rafikov2006}, but see also \citealp{batygin2016}). Instead, they are believed to originate further from their host stars and migrate inward to their present-day locations. Investigating the dynamical history of HAT-P-7 may therefore provide valuable insights into the migration mechanisms of hot Jupiters. For concreteness, we will assume in this study that HAT-P-7b formed in a circular orbit at 3~AU, near the expected location of the ice line \citep{Ida2004}.

\subsection{Simulation setup}\label{ss:setup}
We ran \textit{N}-body simulations using the \texttt{IAS15} integrator \citep{ias15} in \texttt{REBOUND} \citep{rebound}, adopting the system parameters listed in Table~\ref{tab:param}. 

\begin{figure}
    \centering
    \includegraphics[width=1\linewidth]{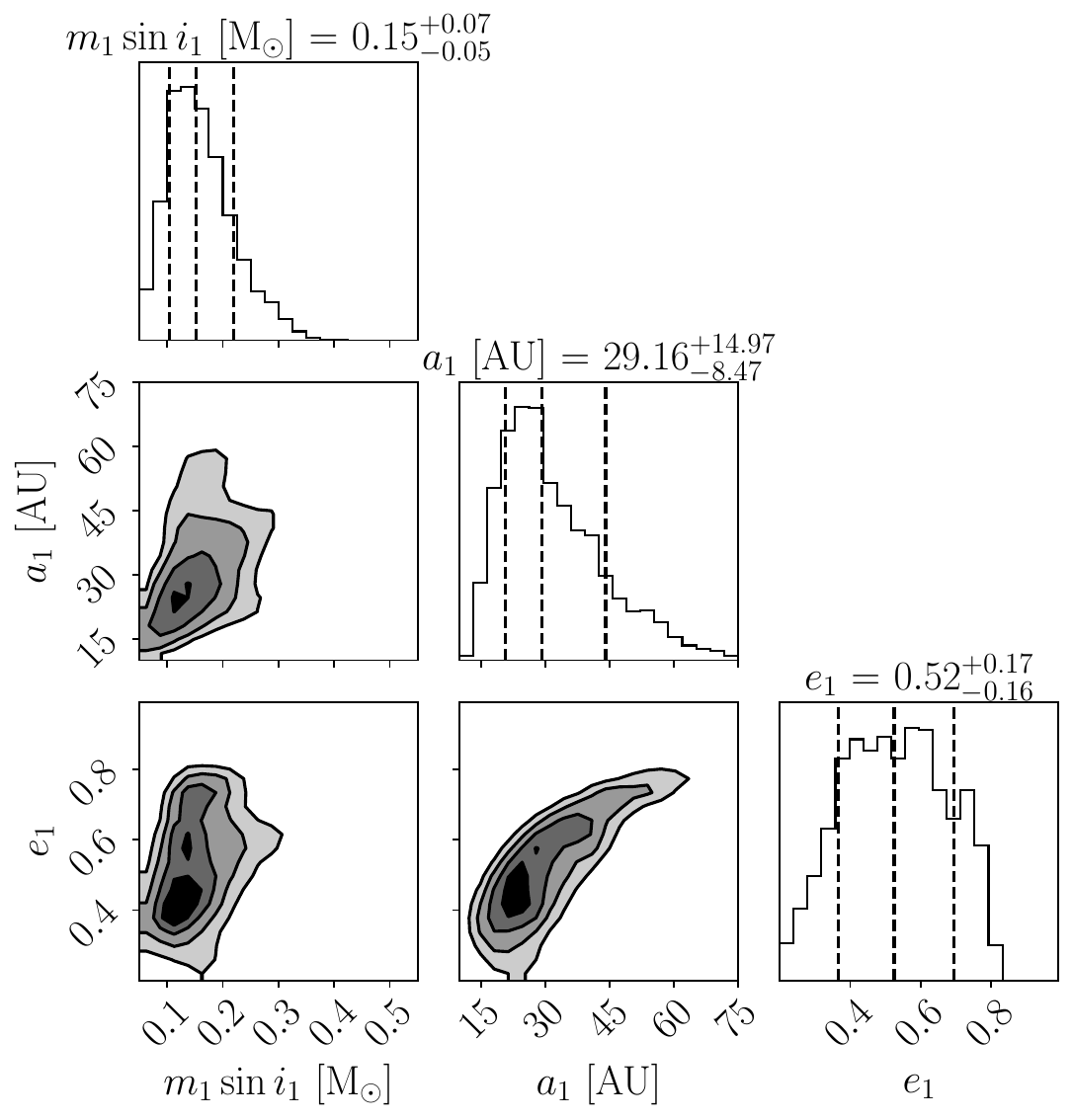}
    \caption{Constraints on the inner companion's orbital parameters, after assuming that the planet formed at 3~AU and applying the long-term stability requirements specified in Equation~\eqref{eq:stability}.}
    \label{fig:param_space_c}
\end{figure}

The posterior distributions shown in Figure~\ref{fig:corner} provide constraints on the orbit of the inner companion. However, not all solutions correspond to physically well-motivated configurations. Long-term stability of the planet against perturbations from the inner companion requires
\begin{equation}
\begin{aligned}
    \frac{a_p}{a_1} \lesssim~ &0.464 - 0.380\mu - 0.631e_1 + 0.586\mu e_1 \\
    &+ 0.150e_1^2 - 0.198\mu e_1^2,\label{eq:stability}
\end{aligned}
\end{equation}
where $\mu \equiv m_1/(M_\star + m_1)$ \citep{Holman1999}.
To enforce this condition, we discarded posterior samples of the inner companion's orbital parameters that are dynamically unstable in the presence of a cold Jupiter at $a_p=3$~AU. The revised posterior distributions are shown in Figure~\ref{fig:param_space_c}.
Accordingly, we adopted a fiducial initial eccentricity of $e_1\simeq0.5$, a semi-major axis of $a_1 \simeq 28$~AU, and a minimum mass of $m_1\sin{i_1} \simeq 0.15\,\rm M_\odot$ for the inner companion. Following a geometric argument outlined in Appendix \ref{app:inc}, we assumed $\sin{i_1}=0.7$, yielding $m_1=0.21\,\rm M_\odot$. Note that if the planet's initial orbital radius were larger than the assumed value of 3~AU, the constraints on the inner companion’s semi-major axis and projected mass would remain similar, while the expected eccentricity would decrease.

Based on its luminosity and color, \cite{Narita2012} reported that the outer companion is an M5 or M5.5 dwarf star. Using Table~5 from \citet{Pecaut2013}, the expected mass
of such a star is about $0.15\,\rm M_\odot$, the value we adopted for modeling purposes. The sky-projected distance between the outer companion and the planet-hosting star is 1240~AU. In general, binaries with separations of
this order of magnitude have a mildly superthermal eccentricity distribution
of
$f_e(e) \propto e^{1.2}$ \citep{Hwang2022}, for which $\langle e\rangle = 0.7$.
Accordingly, we adopted a fiducial eccentricity of $e_2=0.7$ for the outer companion. Since the companion is more likely to be observed near apocenter, we further assumed its semi-major axis to be $a_2 = d_{\rm proj} / (1+e_2) = 730$~AU.

We incorporated the effects of general relativity and tides using \texttt{REBOUNDx} \citep{reboundx}. 
For general relativistic precession, we adopted the \texttt{gr} module \citep{Anderson1975}, appropriate for systems with a dominant central mass. Tidal effects were modeled using the \texttt{tides\_spin} module \citep{Eggleton1998, Lu2023_reboundxspin}, which self-consistently evolves both the spin and orbital angular momenta of bodies under a constant time lag prescription, accounting for tides raised on both the primary and the orbiting companion. This module parameterizes tidal dissipation through a constant time lag $\tau$, which we set to $\tau=1/2nQ$, where $n$ is the mean motion and $Q$ is the tidal dissipation quality factor.

The host star was assumed to have an initial rotational velocity of $15\,\rm km/s$ \citep[e.g.][]{Beyer2024}, a rotation axis aligned with the orbit of the cold Jupiter, a tidal Love number of $0.03$ (similar to that of a solar type star; \citealt{Mecheri2004}), and a tidal dissipation quality factor of $Q_\star=10^7$. 
For the planet, we adopted an initial rotation period of 10~hr, approximately the present-day rotation period of Jupiter. We assumed the planet's tidal Love number to be $k_{2p}=0.5$, similar to Jupiter's. We assumed its tidal quality factor to be $Q_p=10^4$, which is lower than the value inferred for Jupiter by \cite{Goldreich1966}; this choice is discussed further in Section~\ref{ssec:compare}. In the simulations, the planet was assumed to be tidally disrupted whenever its pericenter distance dropped below $2 R_p(M_\star/m_p)^{1/3} = 0.013$~AU \citep{Guillochon2011}.

\begin{table}[]
    \centering
    \caption{Parameters used in \texttt{REBOUND} simulations.}
    \begin{threeparttable}
        \begin{tabularx}{\columnwidth}{l l l}
            \hline
            Parameters & Definition & Values\\
            \hline
            $M_\star$, $m_p$, & Masses & 
            $1.35\,\rm M_\odot$\tnote{a}, $1.84\,\rm M_J$\tnote{a}, \\
            $m_1$, $m_2$ &  & $0.21\,\rm M_\odot$, $0.15\,\rm M_\odot$\\
            $e_p$, $e_1$, $e_2$ & Eccentricities & 0, 0.5, 0.7\\
            $a_p$, $a_1$, $a_2$ & Semi-major axes & 3~AU, 28~AU, 730~AU\\
            $R_\star$, $R_p$ & Radii & $1.99\,\rm R_\odot$\tnote{a}, $16.9\,\rm R_\oplus$\tnote{a}\\
            $k_{2\star}$, $k_{2p}$ & Tidal Love numbers & 0.03, 0.5\\
            $Q_{\star}, Q_p$ & Tidal quality factors & $10^7$, $10^4$\\
            $\Omega_\star$, $\Omega_p$ & Spin frequencies & $2\pi/(7\,\rm d)$, $2\pi/(10\,\rm hr)$\\
            \hline
        \end{tabularx}

        \begin{tablenotes}
            \footnotesize
            \item[a] From \cite{Stassun2017} and \cite{Stassun2019}.
        \end{tablenotes}
    \end{threeparttable}
    \label{tab:param}
\end{table}

Below, we use these parameters and simulations to study two categories of migration pathways: (1) the inner companion directly induces high-e migration of the planet, while the outer companion is dynamically irrelevant; and (2) the outer companion indirectly influences migration by perturbing the inner companion’s orbit, which subsequently triggers the planet's high-e migration.

\subsection{Three-body migration process \label{ssec:3B}}

If the initial mutual inclination between the planet's orbit and the inner companion's orbit around the host star exceeds $\sim 48^\circ$, then the inner companion can readily excite the planet's eccentricity to large values via the octupole-order ZLK mechanism \citep{Naoz2013,Li2014,Munoz2016}. This causes the planet's migration on a timescale that is $\sim\mathcal{O}(10^{-2})$ times the secular timescale of interactions between the inner and outer companions. Because the outer companion is dynamically irrelevant in this scenario, we removed the outer companion from the simulations for simplicity.

\begin{figure}
    \centering
    \includegraphics[width=1\linewidth]{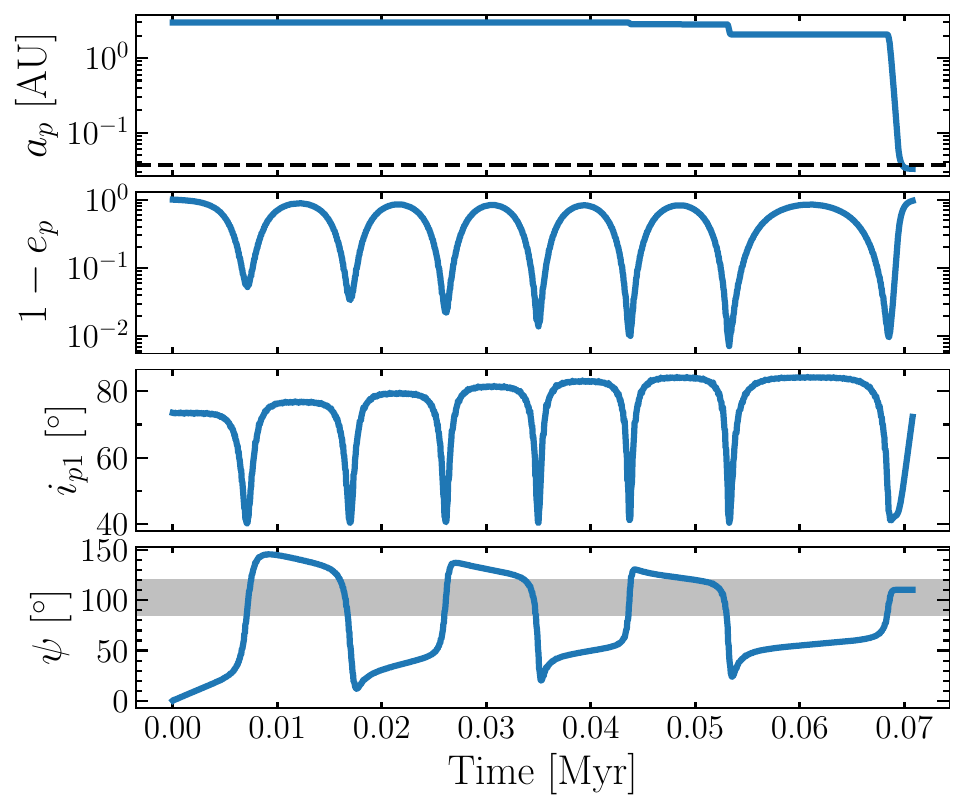}
    \caption{Formation of a retrograde hot Jupiter via the ZLK mechanism induced by the inner stellar companion. The four panels show the time evolution of the planet's semi-major axis, the planet's eccentricity, the mutual inclination between the orbits of the planet and the inner companion, and the spin-orbit angle between the host star's spin axis and the planet's orbital axis. The dashed line in the top panel represents the present-day location of HAT-P-7b. The gray region in the bottom panel indicates the range of spin-orbit misalignment angle of the HAT-P-7 system reported by various studies \citep{Winn2009, Narita2009, Benomar2014, Lund2014, Masuda2015}.}
    \label{fig:example_no_B}
\end{figure}

Figure~\ref{fig:example_no_B} shows an example of the three-body migration process.
The planet begins on a circular orbit with $a=3$~AU (top two panels), and a large inclination relative to the the inner companion (third panel).
Due to this large initial inclination, the ZLK mechanism drives the planet’s eccentricity to extreme values over a few cycles. At peak eccentricity, tidal dissipation becomes rapid enough for the planet to begin migrating inward. The planet eventually circularizes near its present-day location (top panel), and its final orbit is retrograde (bottom panel).

\begin{figure}
    \centering
    \includegraphics[width=1\linewidth]{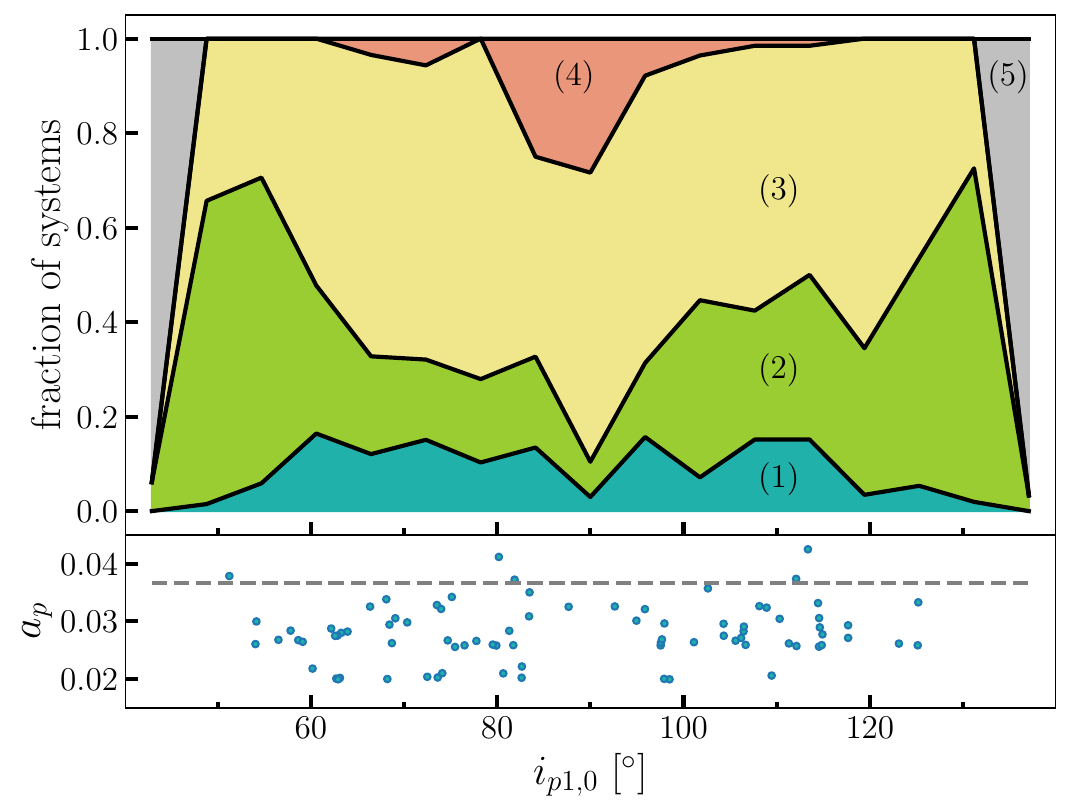}
    \caption{Outcomes of three-body simulations. (Top) Categorization of the state of the system after $50\,\rm Myr$, as a function of the initial mutual inclination between the orbits of the planet and the inner companion. The colors convey the fraction of systems in which the planet (1) becomes a retrograde hot Jupiter, (2) becomes a prograde hot Jupiter, (3) is tidally disrupted, (4) collides with the host star, and (5) does not migrate. The distribution is roughly symmetric about 90$^\circ$. (Bottom) Final semi-major axis of the planet immediately after circularization, for simulations that led to retrograde hot Jupiters. The median value is 0.0275~AU. The dashed line shows the observed value of 0.0367~AU.}
    \label{fig:3B_hist}
\end{figure}

To explore the range of initial configurations that lead to retrograde hot Jupiter formation, we randomly sampled 1000 mutual inclinations from the interval $[40^\circ,140^\circ]$, i.e.\ the ZLK window. 
Figure~\ref{fig:3B_hist} categorizes the states of these systems after $50\,\rm Myr$, a timespan that is much longer
than the octupole-order ZLK timescale of $\sim 0.01\,\rm Myr$ \citep{Antognini2015}.
While retrograde hot Jupiter formation is never the most likely outcome,
its probability is highest when 
the initial mutual inclination is between about
60$^\circ$ and $110^\circ$.

\subsection{Four-body migration process \label{ssec:4B}}

The three-body migration process proposed in the previous section requires a large initial misalignment between the orbits of the planet and the inner companion to activate the ZLK mechanism.
However, such a large misalignment between a planet's orbit and such a close companion is not expected from planet formation theory.
Observations suggest that the orbits of planets are moderately preferentially aligned with those of stellar companions out to separations of $\sim700$~AU \citep{Christian2022, Dupuy2022}.
While this preference disappears when considering only gas giant planets \citep{Christian2024}, the sample is dominated by hot Jupiters, many of which likely underwent subsequent dynamical evolution from their primordial geometries.
In this section, we consider how to induce migration of the cold Jupiter even when it is initially well aligned with the inner companion's orbit, by reintroducing the distant outer companion.

\begin{figure}
    \centering
    \includegraphics[width=1\linewidth]{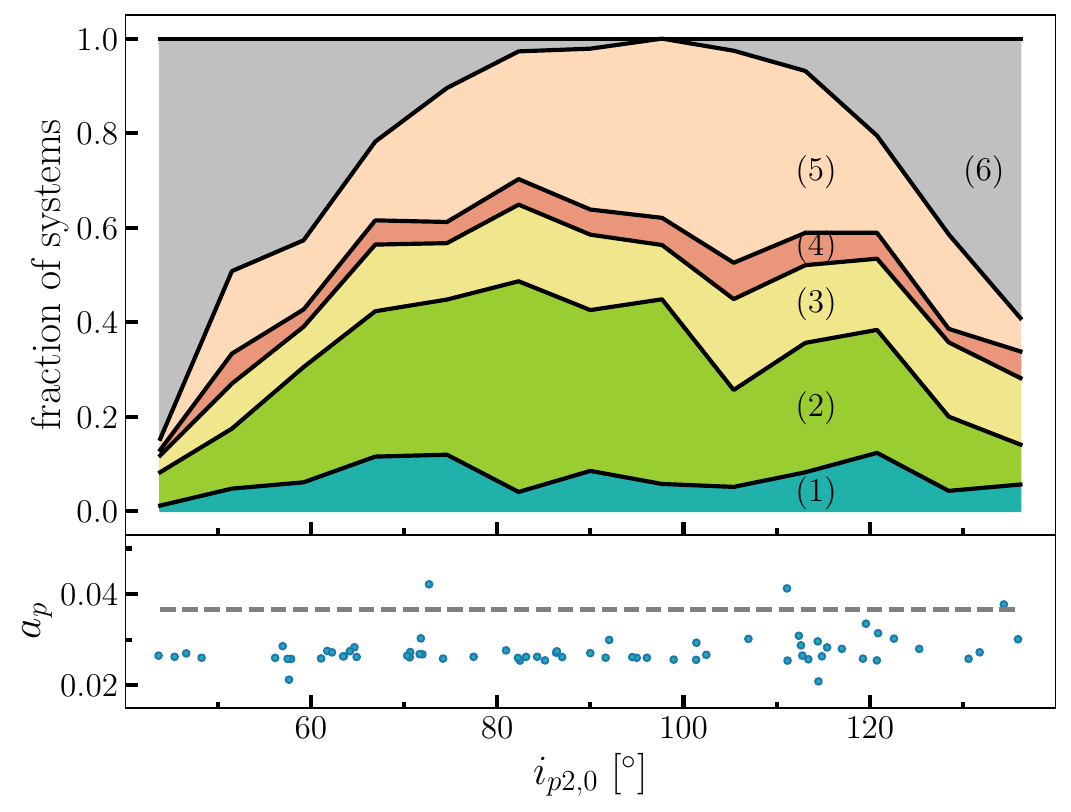}
    \caption{Outcomes of four-body simulations. The initial mutual inclination between the orbits of the inner companion and the planet was fixed at $10^\circ$. (Top) Categorization of the state of the system after 400 Myr, as a function of the initial mutual inclination between the orbits of the
    outer companion and the planet. 
    Colors convey the fraction of systems in which the planet (1) becomes a retrograde hot Jupiter, (2) becomes a prograde hot Jupiter, (3) gets tidally disrupted, (4) collides, (5) escapes the system, and (6) does not migrate. (Bottom) Final semi-major axis immediately after circularization, for simulations that led to retrograde hot Jupiters. The median value is 0.0266~AU. The dashed line shows the observed value of 0.0367~AU.}
    \label{fig:4B_hist}
\end{figure}

\begin{figure}
    \centering
    \includegraphics[width=0.95\linewidth]{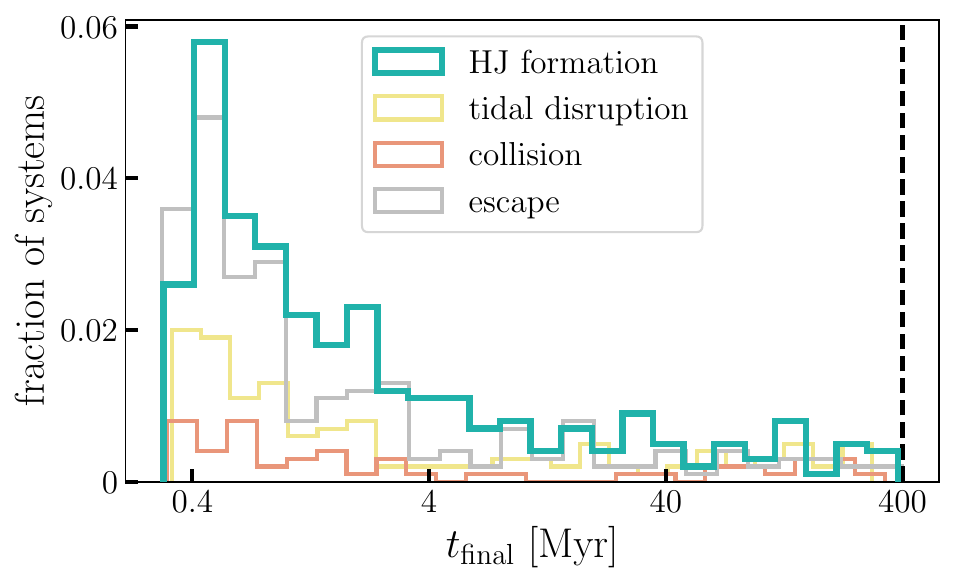}
    \caption{Distribution of times at which the planet becomes a hot Jupiter, is tidally disrupted, collides, or escapes the system. The dashed line marks the end of the simulation runtime (400~Myr). Most hot Jupiter formation occurs within the first few million years.}
    \label{fig:t_final}
\end{figure}

To investigate this scenario, we adopted the parameters listed in Table~\ref{tab:param} and set the initial mutual inclination to be $i_{p1,0}=10^\circ$ between the orbits of the planet and the inner companion. We sampled 1000 values of the mutual inclination between the orbits of the outer companion and the planet drawn from the interval $[40^\circ, 140^\circ]$. Figure~\ref{fig:4B_hist} categorizes the state of the systems at 400~Myr, which is $\sim 10$ times longer than the octupole-order ZLK timescale associated with interactions between the companion stars\footnote{Note that the distribution of outcomes is asymmetric around 90$^\circ$, unlike that shown earlier in
Figure~\ref{fig:3B_hist}. This is because ZLK dynamics are only symmetric about $90^\circ$ when the angular momentum ratio of the inner and outer companions is very small \citep{Anderson2017_kozai, mangipudi2022_kozai}.}.
Figure~\ref{fig:t_final} shows the distribution of the times at which the planet reached its final dynamical state. This shows that the hot Jupiter formation rate tapers off after a few million years. Thus, by truncating the simulations at 400~Myr, we are probably underestimating the probability of hot Jupiter formation by only a modest amount.
We discuss the details of the dynamical processes below.

\subsubsection{Strongly coupled inner pair: eccentricity cascade \label{ssec:4B_coupled}}
\begin{figure*}
    \centering
    \includegraphics[width=0.7\linewidth]{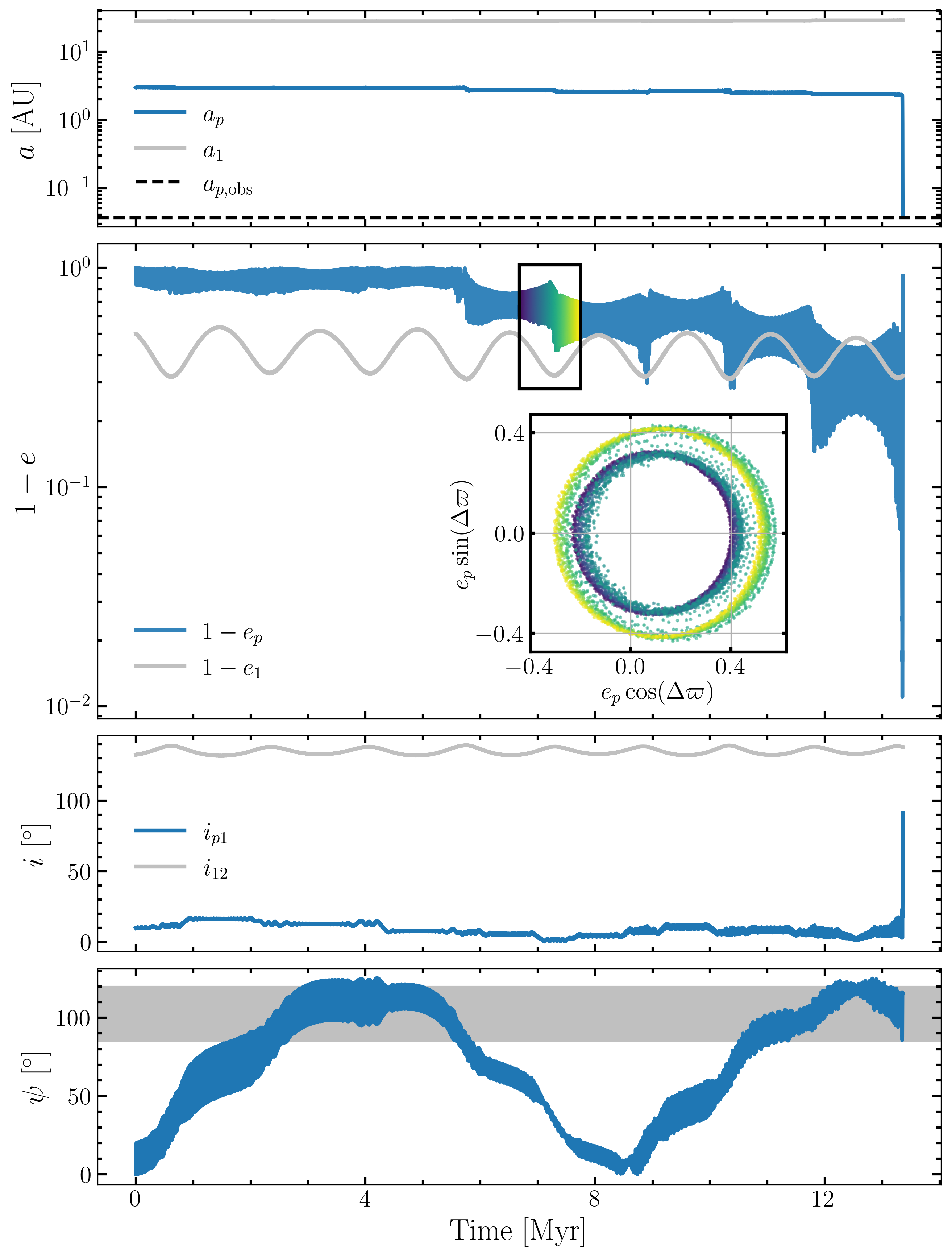}
    \caption{Formation of a retrograde hot Jupiter via the eccentricity cascade migration process. The planet (blue) and the inner companion (gray) initially have a mutual inclination of $10^\circ$, while the inner and outer companions start with a mutual inclination of $138^\circ$. 
    The eccentricity and inclination of the inner companion undergo ZLK oscillations induced by the outer companion. Whenever the inner companion reaches high eccentricity, it undergoes multiple, closely-spaced, weak close encounters with the planet, triggering a sudden increase in the planet’s eccentricity. 
    (Inset) Evolution of the planet’s eccentricity vector during one of these impulsive changes, with color indicating time progression. A close encounter with the inner companion results in an impulsive increase of $\boldsymbol{e}_{\rm free}$, which can be seen as an increase in the radius of the circle traced by the eccentricity vector.}
    \label{fig:example_with_B}
\end{figure*}

For the fiducial parameters, the orbits of the planet and the inner companion remain nearly coplanar, even as the inner companion's orbit varies dramatically due to ZLK oscillations induced by the outer companion. This can be understood by considering the ratio of secular precession frequencies in the system \citep{Lai2017}:\footnote{This equation was derived for the case of circular orbits. For this qualitative argument, we are neglecting the eccentricity dependent terms.}
\begin{equation}
    \frac{\Omega_{12}}{\omega_{p1}} \simeq \frac{m_2}{m_1}\frac{a_1^{9/2}}{a_2^3\, a_p^{3/2}} \sim \mathcal{O}(10^{-3}),~{\rm in~this~case},
\end{equation}
where $\Omega_{12}$ is the orbital precession frequency between the inner and outer companions, and $\omega_{p1}$ is the same between the planet and inner companion. Given the small value of the ratio, we expect
the planet's precession to be strongly coupled to that of the inner companion, maintaining a low mutual inclination.
Thus, the outer companion cannot directly tilt the planet's orbit away from that of the inner companion.
However, the outer companion can excite the inner companion's eccentricity, which in turn excites the planet's eccentricity and triggers high-e migration, as illustrated in Figure~\ref{fig:example_with_B}. We refer to this process as the ``eccentricity cascade'' mechanism.

When the mutual inclination between the two companions' orbits is sufficiently large, the ZLK mechanism can periodically excite the inner companion's eccentricity
and inclination, as
shown by the gray curves in the second and third panels of Figure~\ref{fig:example_with_B}. The ZLK timescale is much longer than the secular timescale of the planet and inner companion. Consequently, the host star, the planet, and the inner companion effectively form a dynamically well-coupled three-body subsystem, where the inner companion's eccentricity is slowly (adiabatically) forced by the ZLK cycles induced by the outer companion.

\begin{figure}
    \centering
    \includegraphics[width=0.5\linewidth]{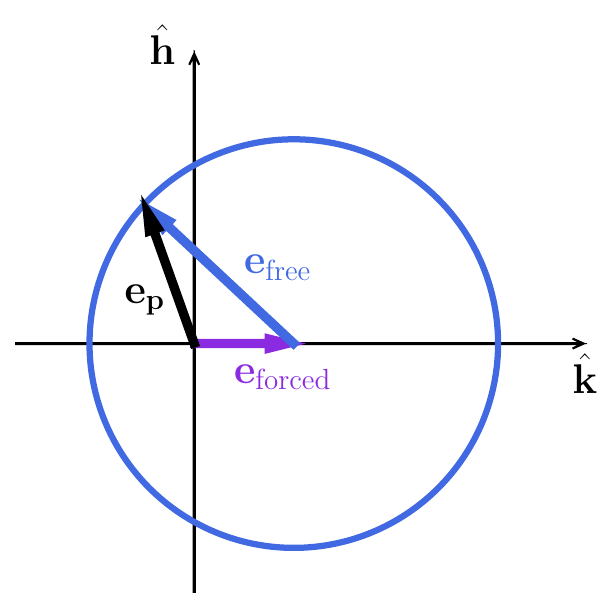}
    \caption{To leading order in eccentricity, the evolution of $\boldsymbol{e}_p$ can be decomposed into a fixed mode $\boldsymbol{e}_{\rm forced}$ and a circulating mode $\boldsymbol{e}_{\rm free}$ in the $\hat{\boldsymbol{k}}$-$\hat{\boldsymbol{h}}$ plane.}
    \label{fig:L-L}
\end{figure}

The blue curve in the second panel of Figure~\ref{fig:example_with_B} reveals two characteristic behaviors in the evolution of the planet's eccentricity. Most of the time, the planet's eccentricity oscillates within a narrow range of values. However, when $e_1$ is near its maximum value, the planet's eccentricity
can abruptly jump. 
This behavior can be interpreted qualitatively
by analogy to the eccentricity evolution under Laplace-Lagrange secular theory, even though such
an approximation is not strictly valid for the large eccentricities seen here\footnote{
Note that a multipolar expansion of the interaction potential in the small parameter $a_p / a_1$ would be more appropriate for the large eccentricities considered here \citep[e.g.][]{Lee2003_secular, Petrovich2015}.
However, we are using the nomenclature from the Laplace-Lagrange solution for a clearer description of the eccentricity evolution.
}. In Laplace-Lagrange theory, the evolution of the planet's eccentricity is understood by decomposing its eccentricity vector
\begin{equation}
    \boldsymbol{e}_p \equiv e_p\cos{\Delta\varpi}\,\hat{\boldsymbol{k}} + e_p\sin{\Delta\varpi}\,\hat{\boldsymbol{h}}
\end{equation}
into the sum of a forced eccentricity induced by the perturbing body, $\boldsymbol{e}_{\rm forced}$, and a freely precessing eccentricity, $\boldsymbol{e}_{\rm free}$ \citep{Murray2000}, where $\Delta\varpi\equiv \varpi_p-\varpi_1$ is the difference in longitudes of pericenter of the planet and inner companion.
This decomposition is shown in Figure~\ref{fig:L-L}. The magnitude and direction of the forced eccentricity are determined by the secular interaction with the perturbing body, such that, in the $\hat{\boldsymbol{k}}$-$\hat{\boldsymbol{h}}$ plane, $\boldsymbol{e}_{\rm forced} \propto e_1\hat{\boldsymbol{k}}$. The free eccentricity
circulates on a secular timescale.

The oscillatory phases of $e_p$ in Figure~\ref{fig:example_with_B} arise from the circulation of $\boldsymbol{e}_{\rm free}$. Over longer timescales, the amplitude
of $e_p$ changes due to variations in
$\boldsymbol{e}_{\rm forced}$ caused by the adiabatic evolution of $e_1$. The abrupt changes in orbital parameters occur when the inner companion is in the high-eccentricity phase of its own secular (ZLK) oscillations, leading to closer encounters with the planet. The resulting weak, non-secular interactions induce rapid changes in $\boldsymbol{e}_{\rm free}$, which is reflected as sudden jumps in $e_p$ (see the inset panel of Figure~\ref{fig:example_with_B}). This process continues until $e_p$ grows sufficiently large for tidal dissipation to become significant, ultimately causing the planet to spiral inward and become a hot Jupiter.

\subsubsection{Decoupled inner pair: two special cases}
For the fiducial parameters adopted in our four-body simulations, most instances of hot Jupiter formation in Figure~\ref{fig:4B_hist} arise from the eccentricity cascade mechanism described in the previous section. In general, the planet and the inner companion remain strongly coupled and well aligned, despite perturbations from the outer companion. However, while exploring a broader range of initial parameters in our simulations, we identified two scenarios in which the planet and the inner companion become dynamically decoupled. Although rare, we report these cases here for completeness.

\begin{figure*}
    \centering
    \includegraphics[width=0.7\linewidth]{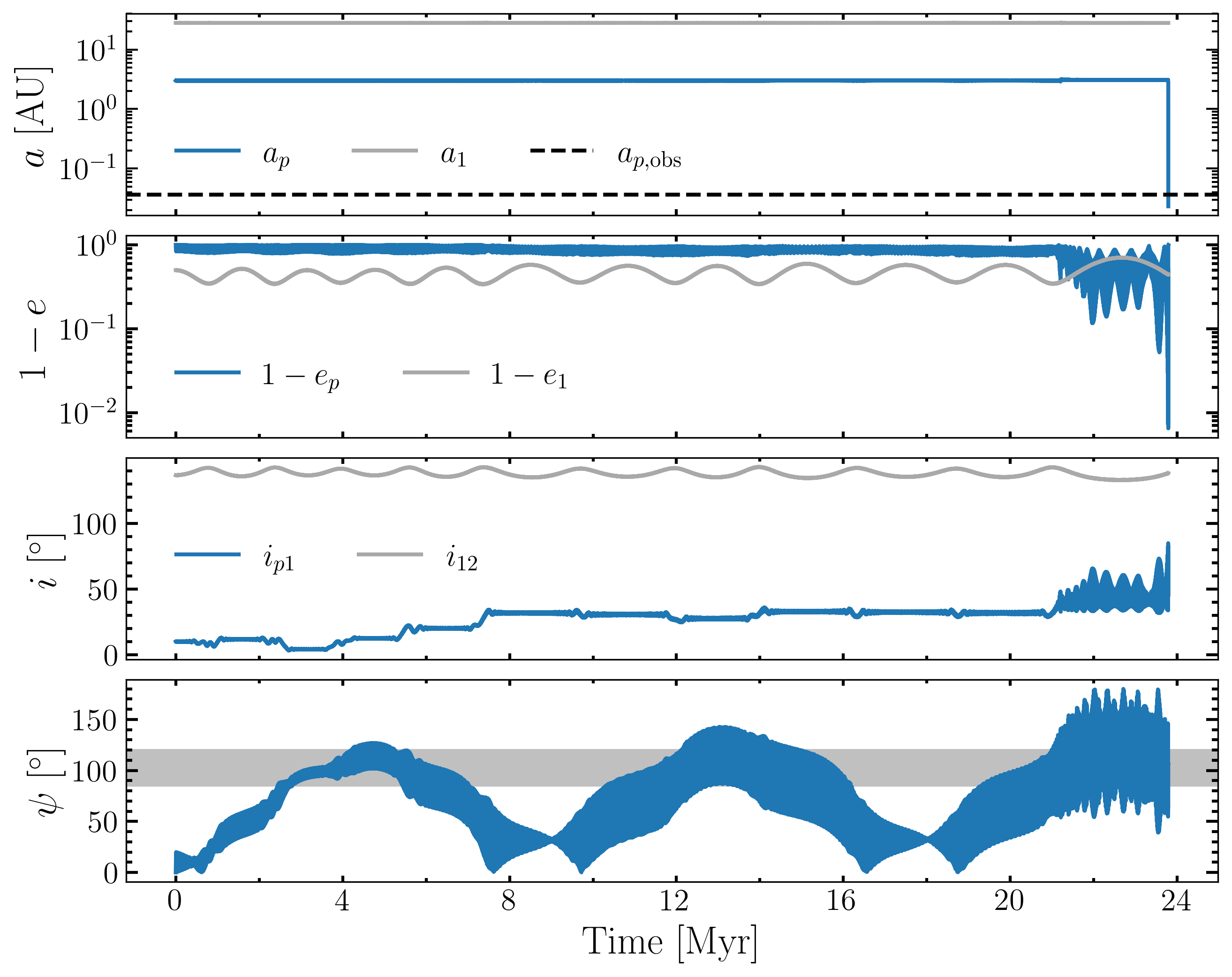}
    \caption{Close encounters between the planet and the inner companion produce impulsive changes in the planet’s inclination. The mutual inclination exceeds the critical angle for ZLK oscillations at $\sim$21~Myr, triggering the planet's high-e migration via the ZLK mechansim.}
    \label{fig:Kozai}
\end{figure*}

The first decoupling pathway is observed in a subset of our four-body simulations. During close encounters, the inner companion induces impulsive changes not only to the planet’s eccentricity but also to its inclination. As illustrated in Figure~\ref{fig:Kozai}, successive inclination kicks can accumulate and gradually increase the mutual inclination between the planet and the inner companion. Once the inclination exceeds the critical threshold for ZLK oscillations, the planet undergoes high-e migration driven by the inner companion. This formation channel is uncommon, as inclination changes from such encounters are typically much smaller than those in eccentricity.

\begin{figure*}
    \centering
    \includegraphics[width=0.7\linewidth]{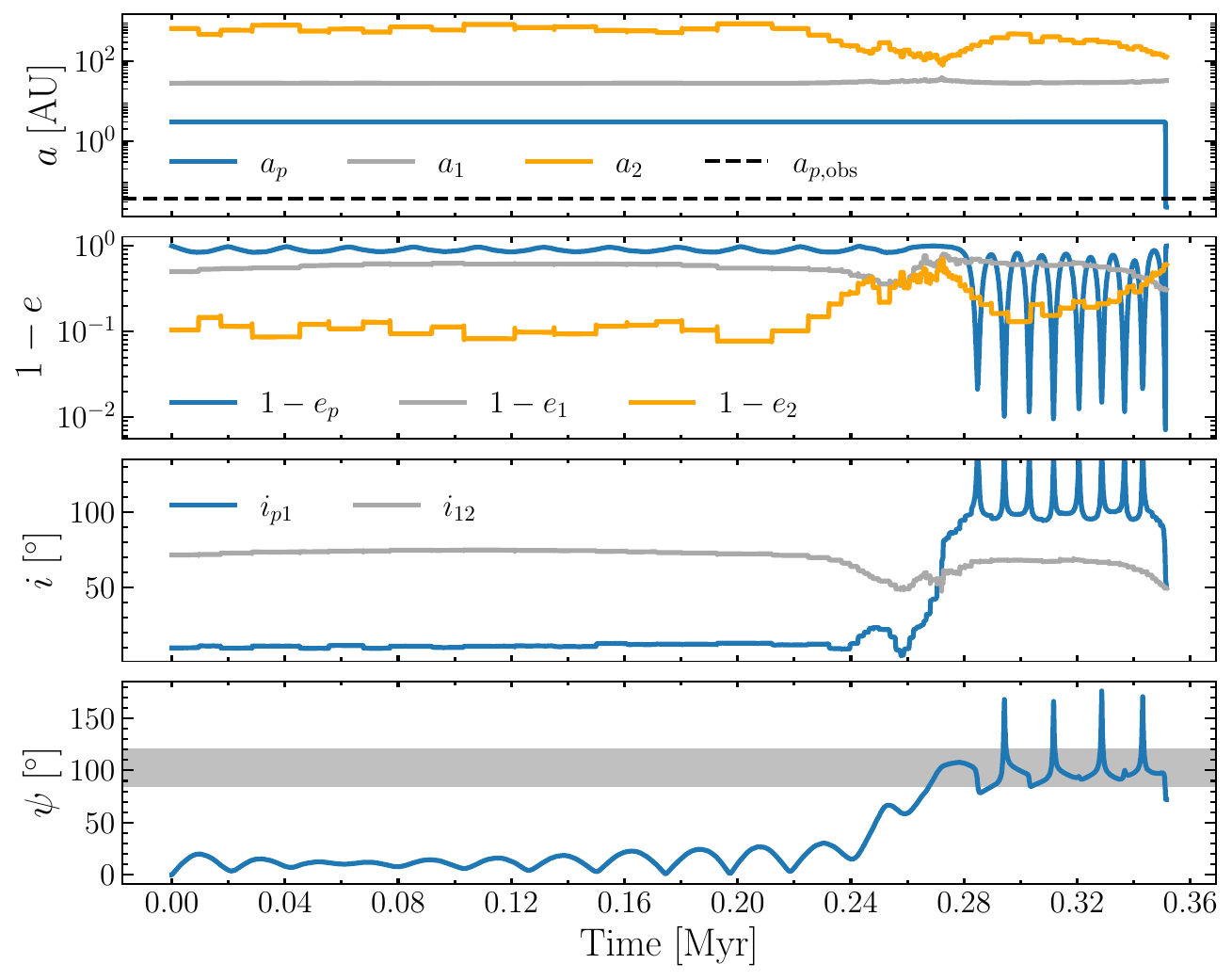}
    \caption{A novel two-step process for ``stimulated'' high-e migration that was observed when the outer companion was assigned an initial eccentricity of 0.9. A close encounter between the inner and outer companions misaligned the planet and the inner companion at a time of $\sim$0.26~Myr. The misalignment exceeded the critical angle for ZLK oscillations, and the planet underwent high-e migration.}
    \label{fig:decouple}
\end{figure*}

The second decoupling scenario appears when adopting a high initial eccentricity for the outer companion ($e_2 \simeq 0.9$). In some of these cases, the outer companion passes very close to the inner companion, impulsively scattering it onto a more inclined orbit. Once misaligned, the inner companion initiates the ZLK mechanism with the planet and triggers high-e migration, as illustrated in Figure~\ref{fig:decouple}. 
Because this scenario requires a rather high initial eccentricity for the outer companion, and because 99\% of the simulations in this scenario resulted in the ejection of either the planet or the outer companion, we consider it to be unlikely as an explanation for HAT-P-7b.

\section{Discussion \label{sec:discussion}}
In this paper, we have shown that the previously reported long-term radial velocity trend of the HAT-P-7 host star (see Figure~\ref{fig:RV}) is caused by an M dwarf star with an orbital distance of several tens of AU and an eccentricity higher than 0.5 (see Figure~\ref{fig:corner}). When combined with the previously identified wide-orbiting stellar companion at $\sim$1000~AU, we now know that HAT-P-7 is a hierarchical triple star system.

We have demonstrated that multiple dynamical mechanisms could have transformed a planet on an initially circular orbit at 3~AU into a hot Jupiter resembling the observed planet HAT-P-7b. If the inner companion's orbit were initially highly misaligned with that of the planet, the inner companion could have initiated ZLK oscillations leading to high-e
migration (see Section~\ref{ssec:3B}, Figures~\ref{fig:example_no_B} and \ref{fig:3B_hist}). Alternatively, and perhaps more plausibly, the planet and the inner companion may have formed in a well-aligned configuration, with the outer companion indirectly triggering migration by perturbing the inner companion’s orbit.

In this scenario, we have proposed an eccentricity cascade mechanism, in which secular modulation of the inner companion's eccentricity, driven by the outer companion, periodically brings the inner companion close enough to the planet to induce impulsive changes in the planet's eccentricity, eventually leading to high-e migration (see Section~\ref{ssec:4B_coupled}, Figure~\ref{fig:example_with_B}). We have shown that this mechanism can plausibly produce retrograde hot Jupiters across a broad range of the outer companion’s inclination space (see Figure~\ref{fig:4B_hist}). More broadly, we have found that even a very distant stellar companion can facilitate hot Jupiter formation by perturbing the orbit of a closer-orbiting companion.

\subsection{Proximity of the inner companion}
The properties of the inner companion star are somewhat unusual. Multiple studies have found that the host stars of hot Jupiters have a lower probability of having close stellar companions than stars without hot Jupiters \citep{Knutson2014, Wang2014, Wang2015a, Wang2015b, Ngo2016, Kraus2016, Ziegler2020}. 
To our knowledge, the only confirmed hot Jupiter system with a stellar companion within 50~AU is WASP-11. In that system, the host star of the hot Jupiter has a low-mass stellar companion
with a projected separation of $42\pm2$~AU \citep{Knutson2014,Ngo2015}. Such close stellar companions are generally thought to inhibit planet formation, as a companion at binary separation $a_1$ would truncate the circumprimary disk to have an outer radius of $\sim 0.3a_1$\citep{Artymowicz1994}. However, observations have shown that giant planet formation is not precluded in compact binary systems, as exemplified by $\gamma$ Cephei \citep{Jang-Condell2015}.

\subsection{Parameter space of the outer companion}
The high eccentricity of the inner companion suggests that the outer companion may play an important dynamical role, potentially exciting the inner companion's eccentricity from an initially low value via the ZLK mechanism. In our \textit{N}-body simulations, we assumed that the outer companion has an eccentricity of $0.7$ and is observed near apocenter.
Had we adopted a lower eccentricity, or a current location away from apocenter, the characteristic timescale for dynamical evolution would have been longer, and the computational cost of the simulations would have been greater. However, as we show in Appendix~\ref{app:eps_rot}, the outer companion can induce ZLK oscillations over a broad range of orbital parameters, and hot Jupiters can still form well within 2~Gyr, the observed age of the system.

\subsection{Comparison with observed orbital properties \label{ssec:compare}}
The final semi-major axes of the retrograde hot Jupiters in our simulations span a range that includes the observed semi-major axis of HAT-P-7b, but the mean of the distribution is systematically smaller by $\sim 25\%$ than the observed value. This discrepancy could potentially be resolved by accounting for radius inflation. During the planet’s inspiral, the dissipation of tidal disturbances within the planet produce internal heat that can inflate the planet to as much as twice its original radius \citep{Bodenheimer2001, Thorngren2021}. Since the tidal precession rate scales as $R_p^5$, an inflated planetary radius would lead to stronger tidal effects, allowing the planet's orbit to become circularized at a larger distance from the star. Consequently, the angular momentum of the planet during the tidal migration is larger than the uninflated case, potentially reconciling the discrepancy between the simulated and observed semi-major axes.

The impact of radius inflation has been investigated for the HAT-P-11 system \citep{Lu2025} and the WASP-107 system \citep{Yu2024}.
In those cases, a model incorporating tidally-driven inflation successfully reproduced the present-day semi-major axes of HAT-P-11b and WASP-107b. This suggests that similar effects could play a role in shaping the final orbital configuration of HAT-P-7b.

Additionally, \cite{Wu2018} and \cite{Vick2019} showed that the effects of chaotic tides lead to a broader distribution of final semi-major axes for hot Jupiters compared to predictions from traditional tidal mechanisms.
In this scenario, the oscillation amplitude of the proto-hot Jupiter's fundamental mode (f-mode) grows diffusively due to forcing from the star on successive pericenter passages. The planet's circularization occurs mainly during the brief intervals of strong dissipation when the mode amplitude becomes large enough for nonlinear effects to be important. Because the mode growth is stochastic, this threshold can be crossed at a larger orbital separation than usual.

Both radius inflation and chaotic tides contribute to enhanced tidal dissipation. To account for this effect in a simplified manner, we adopted a low tidal dissipation quality factor for the planet in our simulations, following a similar approach to that of \citet{Anderson2016}. A more sophisticated treatment of tidal dissipation is beyond the scope of our study, but might allow for more accurate predictions for the final orbital separation.

\subsection{Comparison to other systems}
To our knowledge, HAT-P-7 is one of very few systems in which there are plausible dynamical pathways to forming a hot Jupiter without fine tuning and based only on the observed properties of known bodies, as opposed to invoking hypothetical bodies that escaped the system or have properties that make them unobservable. Two other such systems are HD~80606 and TIC~241249530:
\begin{itemize}
    \item The HD~80606 system is a textbook example of the high-e tidal migration pathway. It hosts a giant planet, HD 80606b ($m_p=4.1\,\rm M_J$, $a_p=0.46$~AU, $e_p=0.93$; \citealp{Naef2001}), which is thought to be currently undergoing high-e migration. The presence of a main-sequence companion, HD 80607, at a projected separation of $\sim 1000$~AU suggests the possibility of ZLK migration. \cite{Wu2003} demonstrated that if the planet initially formed at 5~AU, the Kozai mechanism requires an initial mutual inclination of $\sim 85^\circ$--$95^\circ$ between the planet and the distant companion.
    \item TIC~241249530 hosts a hot Jupiter progenitor ($m_p=5.0$~$\rm M_J$, $a_p=0.64$~AU, $e_p=0.94$; \citealp{Gupta2024}) that closely resembles HD 80606. This system has a low-mass stellar companion, TIC 241249532, at a projected separation of 1664~AU. \cite{Gupta2024} showed that the present-day parameters of the system can be reproduced if the planet initially formed beyond 7~AU and the binary companion had a mutual inclination of $ 86.8^\circ$--$93.2^\circ$.
\end{itemize}

Both cases above require the companion’s orbit to be nearly perpendicular to that of the planet. In neither case has an additional stellar companion been detected, but based on our study, it seems worth conducting observations that could test for the presence of an inner companion analogous to 
the inner companion of HAT-P-7.
This could allow for a four-body migration process, loosening the strict inclination constraints imposed by the traditional ZLK migration framework.

Systems with evidence for high-e migration for other types of planets have also been studied in detail:
\begin{itemize}
    \item WASP-107 hosts a Neptune-mass planet, WASP-107b, in a 5.7-day, polar orbit ($m_b=30.5$~$\rm M_\oplus$, $a_b=0.14$~AU, $e_b=0.06$; \citealp{Anderson2017,Dai2017,Piaulet2021, Rubenzahl2021}). The system also has a long-period companion, WASP-107c ($m_c\sin_i \simeq 0.36$~$\rm M_J$, $P_c\simeq2.98$~yr, $e_c\simeq0.28$; \citealp{Piaulet2021}). Simulations suggested that WASP-107b most likely formed within 0.5~AU of the host star, which seems consistent with the observed low C/O value \citep{Welbanks2024}. From such a starting configuration, WASP-107c could have driven octupole-order ZLK oscillations and prompted high-e migration of WASP-107b. This mechanism requires an initial mutual inclination between 50$^\circ$ and 70$^\circ$ between the orbits of the two planets.
    \item Another example is the HAT-P-11 system, which has two known planets. The inner planet, HAT-P-11b, is a close-in, eccentric super-Neptune ($m_b=23.4$~$\rm M_\oplus$, $a_b=0.0525$~AU, $e_b=0.218$; \citealp{Bakos2010}), while the outer planet, HAT-P-11c, is an eccentric super-Jupiter ($m_c\sin_i \simeq 1.6$~$\rm M_J$, $a_c\simeq4.1$~AU, $e_c\simeq0.6$). \cite{Lu2025} proposed that a long-ago epoch of planet-planet scattering led to the ejection of two additional planets and generated a large mutual inclination between HAT-P-11b and HAT-P-11c -- which, in turn, initiated ZLK-driven migration.
\end{itemize}

The scenarios described above for
WASP-107b and HAT-P-11b do not require initial mutual inclinations extremely close to $90^\circ$. The more modest mutual inclinations that are required for these scenarios can be naturally produced through planet-planet scattering. However, these scenarios necessarily involve unobservable bodies that were involved in past scattering events but were ejected from the system. While ejections are certainly possible, this feature makes such scenarios harder to test. Based on our study of HAT-P-7, we wonder whether there could be additional companions in the WASP-107 and HAT-P-11
systems that facilitated high-e migration, without requiring finely tuned initial conditions or ejected bodies.

Proto-hot Jupiters on highly eccentric orbits, such as TOI-3362b \citep{Dong2021} and HAT-P-2b \citep{Bakos2007, Lewis2013, deBeurs2023}, provide additional opportunities to explore few-body migration pathways. Further observational constraints on potential companions in these systems could offer deeper insights into their dynamical histories. Extending radial velocity monitoring baselines and leveraging astrometric surveys may reveal new multi-body configurations that challenge or refine existing high-e migration models.

Although not directly related to hot Jupiter formation, \citet{Best2022} proposed a mechanism similar to the eccentricity cascade to explain the retrograde spin of the host star in the K2-290 system. In the K2-290 system, the host star is the primary star of a close binary, and there is also a distant
stellar companion. \citet{Best2022} proposed that the distant companion excited ZLK oscillations of the close binary. When the close binary was in the high eccentricity phase of the oscillations, the secondary star perturbed the nodal precession of the two planetary orbits around the primary star, leading to resonance crossings between the stellar spin and the planets’ secular modes. This triggered chaos on secular timescales, ultimately causing the host star's obliquity to become retrograde.
The dynamical mechanisms considered in our works are similar in that a companion with an intermediate orbital separation couples a distant stellar companion to the evolution of an inner planetary system.
They are also dynamically distinct as the intermediate companion in their work interacts with the inner planetary system via secular interactions, while our intermediate companion introduces non-secular evolution.

\section*{Acknowledgements}
We thank the anonymous reviewer for their comprehensive feedback, as well as Caleb Lammers, Tiger Lu, and Sarah Millholland for useful discussions.
Some of the data presented herein were obtained at Keck Observatory, a private 501(c)3 non-profit organization operated as a scientific partnership among the California Institute of Technology, the University of California, and the National Aeronautics and Space Administration. The Observatory was made possible by the generous financial support of the W. M. Keck Foundation. The authors wish to recognize and acknowledge the very significant cultural role and reverence that the summit of Maunakea has always had within the Native Hawaiian community. We are most fortunate to have the opportunity to conduct observations from this mountain. We are also grateful to the
members of the California Planet Search team for organizing and executing Keck radial-velocity observations and facilitating long-term monitoring of stars, including HAT-P-7. Work by JNW was partly supported by a NASA Keck PI Data Award, administered by
the NASA Exoplanet Science Institute. 
YS is supported by a Lyman Spitzer, Jr. Postdoctoral Fellowship at Princeton
University.

\appendix
\section{Inclination distribution \label{app:inc}}
In Section~\ref{ssec:joint_analysis}, we constrained the projected mass $m_1\sin{i_1}$ for the inner companion. Typically, the true mass $m_1$ is estimated by assuming an isotropic distribution of the line-of-sight inclination, yielding an expectation value of $\langle \sin{i_1}\rangle=\pi/4$. In our case, however, there are two geometric constraints that, taken together, break the assumption of isotropy: (i) the inner companion must be misaligned with the planet's orbit by more than $\sim 48^\circ$ for the octupole-order ZLK mechanism to be effective, and (ii) the planet’s orbit normal is nearly perpendicular to the line of sight. In this Appendix, we derive the inclination distribution under these constraints and justify the adoption of $\langle \sin{i_1}\rangle=0.7$ in Section~\ref{ss:setup}.

\begin{figure}
    \centering
    \includegraphics[width=0.7\columnwidth]{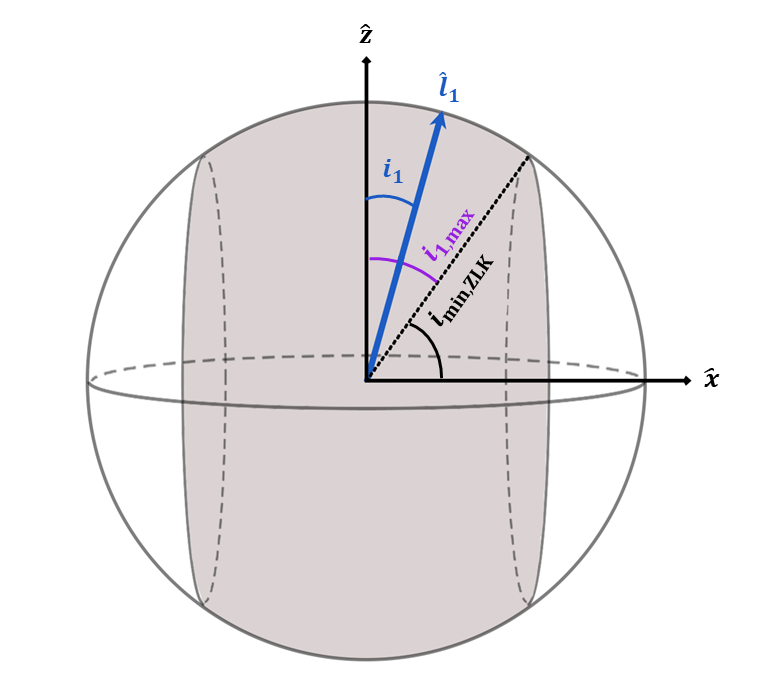}
    \caption{
    Illustration of the coordinate system adopted in Appendix~\ref{app:inc}.
    The line of sight is oriented along $\hatb{z}$ and the planet's orbit normal is along $\hatb{x}$.
    The inner companion's orbit normal $\hatb{l}_1$ must be oriented such that the mutual inclination between the planet and companion exceeds $i_{\rm \min,ZLK} \approx 48^\circ$ for the octupole-order ZLK mechanism to be active.
    The permitted region for $\hatb{l}_1$ is the three-dimensional surface $S$ as defined in Equation~\eqref{eq:restrictedsphere}.
    The objective is to determine the probability distribution of $i_1$ over $S$.
    }
    \label{fig:coords_app}
\end{figure}

We define a coordinate system as shown in Figure~\ref{fig:coords_app}, where $\hatb{z}$ is aligned with the line of sight and the planet's orbit normal is assumed to lie in the $\hatb{x}$-$\hatb{y}$ plane. The inner companion's orbit must have an inclination of at least $i_{\rm min, ZLK}=48^\circ$ (and at most $180^\circ - i_{\rm min, ZLK}$) relative to the planet's orbital plane to induce high-e migration. This constraint restricts the orientation of the companion's orbit normal, $\hatb{l}_1$, to the surface given by
\begin{equation}
    S = \left\{(x, y, z): x^2 + y^2 + z^2 = 1, |x| < \cos i_{\rm \min,ZLK}\right\}.\label{eq:restrictedsphere}
\end{equation}

\begin{figure}
    \centering
    \includegraphics[width=0.7\columnwidth]{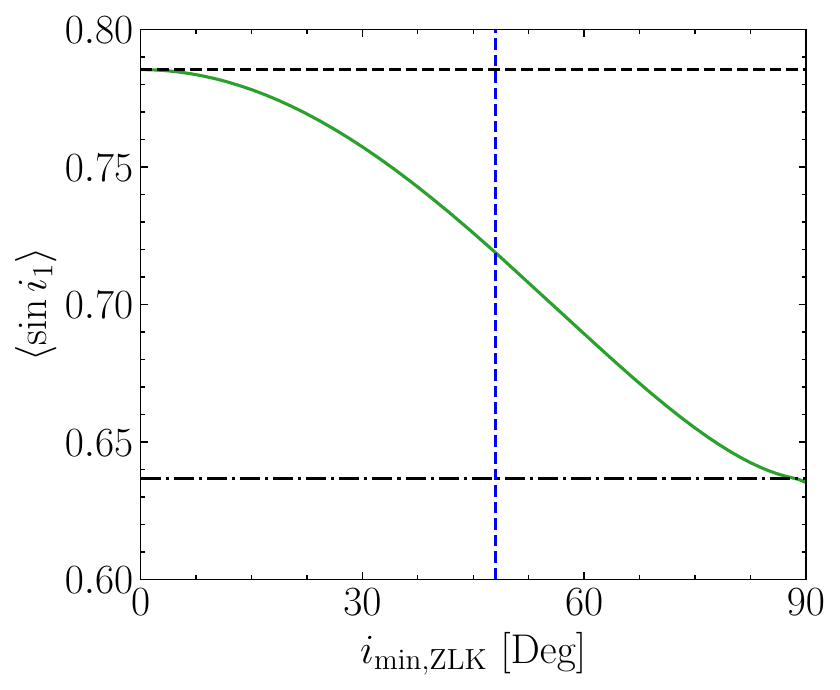}
    \caption{
    Dependence of $\langle \sin i_1 \rangle$ on $i_{\rm \min,ZLK}$.
    The fiducial value of $i_{\rm \min,ZLK} = 48^\circ$ is denoted with the vertical blue dashed line, and the two asymptotic limits of $i_{\rm \min,ZLK} = 0^\circ$ and$i_{\rm \min,ZLK} = 90^\circ$ are shown in the horizontal black dashed and dash-dotted lines, respectively.
    }
    \label{fig:sini_dist}
\end{figure}

To evaluate $\langle \sin i_1\rangle$ over $S$, we first show that, by decomposing the surface along the $z$ axis, the probability density function of $\cos i_1$ is given by
\begin{equation}
    f(\cos i_1) \propto
        \begin{cases}
            2\pi & \sin i_1 < \sin i_{1,\max},\\
            4 \arcsin\left(\frac{\sin i_{1,\max}}{\sin i_1}\right)
                & \sin i_1 > \sin i_{1,\max}.
        \end{cases}\label{eq:pdf_z}
\end{equation}
Then, this distribution can be used to evaluate $\langle \sin i_1\rangle = \langle 1 - \cos^2 i_1\rangle$ numerically.
The mean value as a function of $i_{\rm \min,ZLK}$ is shown in Fig.~\ref{fig:sini_dist}.
We note two useful asymptotic limits.
In the isotropic case where $i_{\rm \min,ZLK}=0^\circ$, the well-known result $\langle \sin i_1 \rangle = \pi/4$ is recovered. If instead $i_{\rm \min,ZLK}\to 90^\circ$, the allowed orientations collapse to a unit circle in the $\hatb{y}$-$\hatb{z}$ plane, yielding $\langle \sin i_1 \rangle = 2/\pi$.
For the fiducial case of $i_{\rm min,ZLK} = 48^\circ$, we find $\langle \sin i_1 \rangle \approx 0.72$.

Strictly speaking, the result provided here should not be applied to the inclinations between the \textit{present-day} hot Jupiter and inner companion for two reasons.
First, our calculation is based on the mutual inclination between the companion and the \textit{proto}-hot Jupiter, which is not equal to the present-day misalignment angle.
This is because the planet's orbital orientation will change during and after its migration due to precession about the host star's spin axis (driven by its rotational bulge).
However, the host star's small degree of rotational line broadening suggests that its spin axis is quite well-aligned with the line of sight (e.g.\ $\sin i_\star \lesssim 1/3$, \citealp{Winn2009}).
As such, the geometry depicted in Fig.~\ref{fig:sini_dist} is likely still accurate up to rotations about the $\hat{z}$ axis, which does not affect the $\sin i_1$ distribution.
Second, not all initial mutual inclinations are equally conducive to hot Jupiter formation; as shown in Figures~\ref{fig:3B_hist} and~\ref{fig:4B_hist}, extreme misalignments are generally disfavored due to the efficiency of tidal disruption.
Nevertheless, a bias towards larger $m_1$ values than those expected from assuming isotropy remains physically well-motivated.
Based on these considerations, we adopt $\langle \sin i_1 \rangle = 0.7$ as a representative value.
We find that small variations in $m_1$ do not significantly affect our conclusions.

\section{ZLK oscillations between the two stellar companions\label{app:eps_rot}}
The Kozai timescale between the inner and outer companions is \citep{Antognini2015}
\begin{equation}
    t_{\rm ZLK} \simeq \frac{16}{15}\left(\frac{a_2^3}{a_1^{3/2}}\right)\sqrt{\frac{M_\star+m_1}{Gm_2^2}} (1-e_2^2)^{3/2}.
\end{equation}
However, ZLK interactions between them can be suppressed by short-range effects that induce pericenter precession in the inner companion. The dominant contribution comes from the quadrupole potential of the planet's orbit, whose secular Hamiltonian is given by
\begin{equation}
    \langle{\Phi_{\rm orb}}\rangle =  - \frac{G m_1[C-A]_p}{2 a_1^3 (1-e_1^2)^{3/2}},
\end{equation}
\\
\noindent where the orbit is modeled as a ring, and the difference in polar and equatorial moments of inertia is $[C-A]_p = \frac{1}{2}m_pa_p^2$. 
The corresponding apsidal precession rate is
\begin{equation}
    \dot{\omega}_{\rm orb} = -\frac{\sqrt{1-e_1^2}}{e_1L_1}\frac{\partial \langle\Phi_{\rm orb}\rangle}{\partial e_1}, \label{eq:omega_orb}
\end{equation}
where $L_1 = M_\star m_1 \sqrt{G a_1 / (M_\star + m_1)}$ is the angular momentum of the inner companion \citep{Fabrycky2007}. 

Following the notation in \cite{Liu2015}, we define a characteristic ZLK rate as
\begin{equation}
    \dot{\omega}_{\rm ZLK} \equiv \frac{1}{t_{\rm ZLK}\sqrt{1-e_1^2}}.
\end{equation}

The ratio of the precession rate due to the planet's orbit from Equation~\eqref{eq:omega_orb} to the characteristic ZLK rate is
\begin{equation}
    \frac{\dot{\omega}_{\rm orb}}{\dot{\omega}_{\rm ZLK}}
    = \frac{3}{4} \frac{(M_\star+m_1)m_p}{M_\star m_2}\frac{a_p^2a_2^3}{a_1^5} \left(\frac{1-e_2^2}{1-e_1^2}\right)^{3/2}. \label{eq:eps}
\end{equation}

We define a dimensionless parameter
\begin{equation}
    \epsilon \equiv \frac{3}{4} \frac{(M_\star+m_1)m_p}{M_\star m_2}\frac{a_p^2a_2^3}{a_1^5} (1-e_2^2)^{3/2}
\end{equation}
such that $\dot{\omega}_{\rm orb}/\dot{\omega}_{\rm ZLK} = \epsilon (1-e_1^2)^{-3/2}$.
The maximum eccentricity that the inner companion can reach is determined by
\begin{equation}
\begin{aligned}
    &\frac{\epsilon}{3}\left(\frac{1}{(1-e_{1,\max}^2)^{3/2}}-1\right) \\
    &= \frac{9}{8}\frac{e_{1,\max}^2}{1-e_{1,\max}^2}
    \left(1-e_{1,\max}^2-\frac{5}{3}\cos^2{i_{12,0}}\right),
\end{aligned}
\end{equation}
where $i_{12,0}$ is the initial mutual inclination between the inner and outer companions \citep{Liu2015}.

\begin{figure}
    \centering
    \includegraphics[width=0.7\columnwidth]{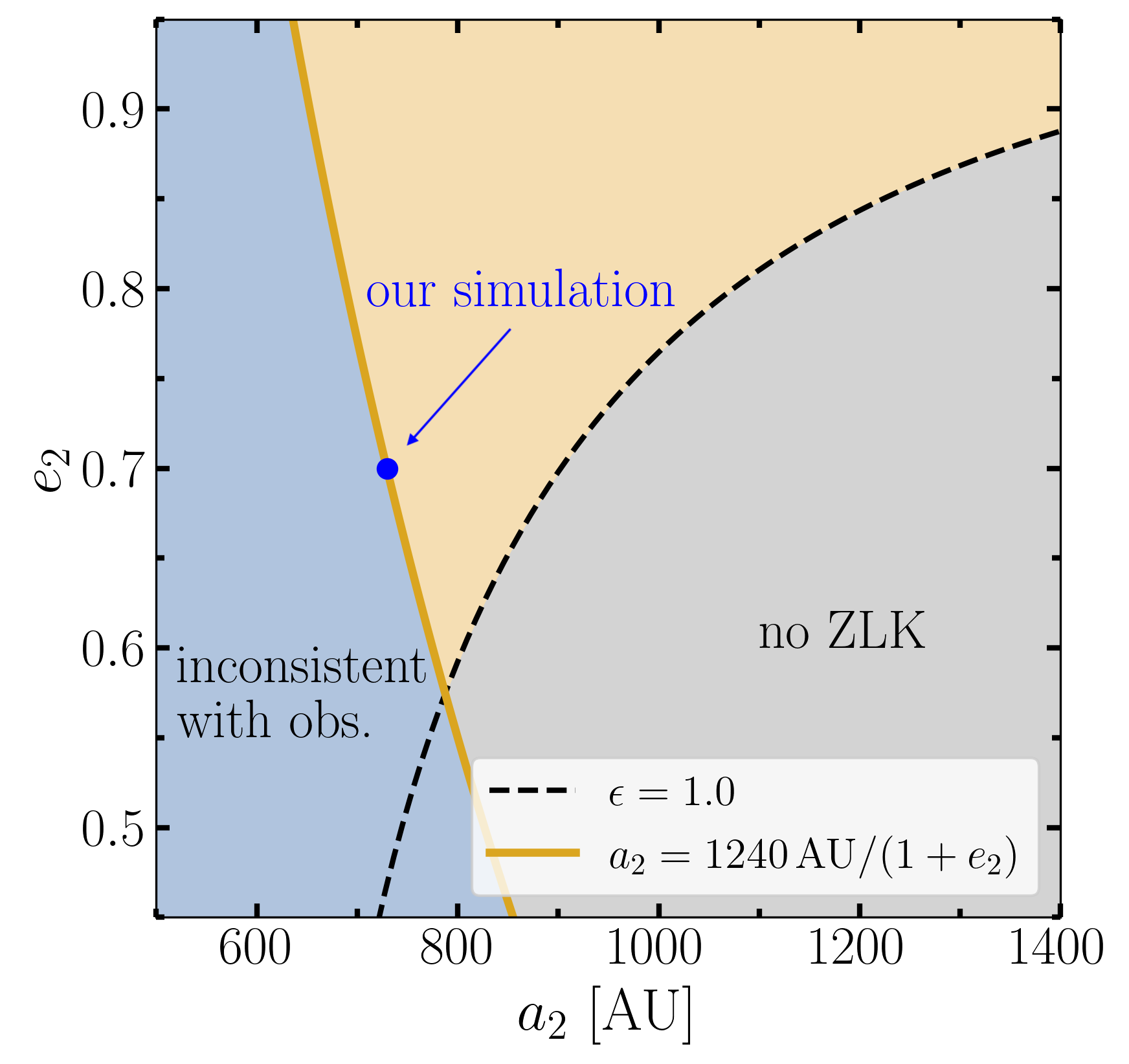}
    \caption{Suppression of ZLK oscillations due to the quadrupole moment of the planet's orbit, as a function of the semi-major axis and eccentricity of the outer companion. The blue dot shows the configuration we adopt in our simulations. Any configurations to the left of the orange line will be inconsistent with the observed projected separation of the outer companion. The inner companion cannot reach the limiting eccentricity required for the eccentricity cascade migration process if $\epsilon$ defined in Equation~\eqref{eq:eps} exceeds $1.0$, indicated by the regions shaded in gray.}
    \label{fig:eps_rot}
\end{figure}

For the eccentricity cascade migration process discussed in Section~\ref{ssec:4B_coupled} to be viable, the inner companion must reach a limiting eccentricity of $e_{1,\max} \simeq 0.7$. To the octupole order of the Kozai mechanism (equivalent to setting $i_{12,0}=90^\circ$), this requirement corresponds to $\epsilon \simeq 1.0$. In Figure~\ref{fig:eps_rot}, we show the parameter space of the outer companion that allows the inner companion to achieve this limiting value via the ZLK mechanism.

\bibliography{Bib}
\bibliographystyle{aasjournal}
\end{document}